\documentclass[reprint,english,aps,floatfix,showpacs,superscriptaddress]{revtex4-1}
\usepackage[T1]{fontenc}
\usepackage[utf8]{inputenc}
\usepackage{amsmath}
\usepackage{graphicx}
\usepackage{amssymb}
\usepackage{color}
\usepackage{hyperref}
\usepackage{verbatim}
\usepackage{natbib}
\usepackage{multirow}

\newcommand{\etal}{\textit{et al}.}

\usepackage{babel}

\begin{document}

\title{Assembly of open clusters of colloidal dumbbells via droplet evaporation}

\author{Hai Pham Van}
\affiliation{Theoretische Physik II, Physikalisches Institut, Universit\"at Bayreuth, Universit\"atsstra{\ss}e 30, D-95440 Bayreuth, Germany}
\affiliation{Department of Physics, Hanoi National University of Education, 136 Xuanthuy, Hanoi, Vietnam}

\author{Andrea Fortini}
\affiliation{Theoretische Physik II, Physikalisches Institut, Universit\"at Bayreuth, Universit\"atsstra{\ss}e 30, D-95440 Bayreuth, Germany}
\affiliation{Department of Physics, University of Surrey, Guildford GU2 7XH, United Kingdom}

\author{Matthias Schmidt}
\email{matthias.schmidt@uni-bayreuth.de}
\affiliation{Theoretische Physik II, Physikalisches Institut, Universit\"at Bayreuth, Universit\"atsstra{\ss}e 30, D-95440 Bayreuth, Germany}


\begin{abstract}
We investigate the behavior of a mixture of asymmetric colloidal dumbbells and emulsion droplets by means of kinetic Monte Carlo simulations.  
The evaporation of the droplets and the competition between droplet-colloid attraction and colloid-colloid interactions lead to the formation of clusters built up of colloid aggregates with both closed and open structures. 
We find that stable packings and hence complex colloidal structures can be obtained by changing the relative size of the colloidal spheres and/or their interfacial tension with the droplets. 
 \end{abstract}

\maketitle
\section{Introduction}
\label{intr}
Complex colloids characterized by heterogeneous surface properties are an active field of research due to their diverse potential applications as interface stabilizers, catalysts, and building blocks for nanostructured materials. 
Janus particles are colloidal spheres with different properties on the two hemispheres. 
They have recently attracted significant attention due to their novel morphologies~\cite{Pawar2010}. 
Corresponding dumbbells consist of two colloidal spheres with different sizes or dissimilar materials~\cite{Claudia2013}. 
Many studies have investigated the self-assembly of colloidal dumbbells into more complex structures, including micelles, vesicles~\cite{Sciortino2009,Liang2008}, bilayers~\cite{Munao2013,Whitelam2010,Avvisati2015} and dumbbell crystals~ \cite{Mock2007,Marechal2008,Ahmet2010}. 
Particularly, open clusters of colloidal dumbbells with syndiotactic, chiral~\cite{Zerrouki2008,Bo2013} and stringlike structures~\cite{Smallenburg2012} are significant because they can be regarded as colloidal molecules~\cite{Blaaderen2003,Duguet2011} that exhibit unique magnetic, optical and rheological properties~\cite{Edwards2007}. 
However, the control of the cluster stability and the particular geometric structure are two major challenges, which have yet to be solved. 
       
Several self-assembly techniques have been used to control the aggregation of colloidal particles. Velev \etal\ developed a method to obtain so-called colloidosomes from colloidal particles by evaporating droplets~\cite{Velev1996a,Velev1996b,Velev1997}.  
Based on this technique, Manoharan \etal ~\cite{Manoharan2003} successfully prepared micrometer-sized clusters and found that the structures of particle packings seem to minimize the second moment of the mass distribution. 
Wittemann \etal\ also produced clusters, but with a considerably smaller size of about 200nm~\cite{Wittemann2008,Wittemann2009,Wittemann2010}.  
Cho \etal\ prepared binary clusters with different sizes or species from phase-inverted water-in-oil~\cite{Cho2005} and oil-in-water emulsion droplets~\cite{Cho2008}. These authors found that the interparticle interaction and the wettability of the constituent spheres play an important role in the surface coverage of the smaller particles. In addition, for oil-in-water emulsions the minimization of the second moment of the mass distribution ($M2$) only applies if the size ratio is less than 3. 
More recently, Peng \etal~\cite{Bo2013} reported both experimental and simulation work on the cluster formation of dumbbell-shaped colloids. These authors proved that the minimization of the second moment of the mass distribution is not generally true for anisotropic colloidal dumbbell self-assembly. However, they predicted cluster structures without considering the different wettabilities for constituent colloidal spheres. 

In previous work, Schwarz \etal~\cite{Ingmar2011} studied cluster formation via droplet evaporation using Monte Carlo (MC) simulation with shrinking  droplets. It was shown that a short-ranged attraction between colloidal particles can produce $M2$ nonminimal isomers and the fraction of isomers varied for each number of constituent particles. 
In addition, supercluster structures were found with complex morphologies starting from a mixture of tetrahedral clusters and droplets. 
Fortini~\cite{Fortini2012} modeled cluster formation in hard sphere-droplet mixtures, without shrinking droplets, and observed a transition from clusters to a percolated network that is in good agreement with experimental results.   

In the current paper, we extend the model of Ref.~\cite{Ingmar2011} in order to investigate the dynamic pathways of cluster formation in a mixture of colloidal dumbbells and emulsion droplets. 
By varying the size or hydrophilic property of colloidal dumbbells, we find a variety of complex cluster structures that have not been observed in clusters of monodispersed colloidal spheres. In particular, we find open clusters with  a compact core, which determines the overall symmetry, and protruding arms. These structures could lead to novel self-assembled structures.

This paper is organized as follows.  We introduce the model and simulation method in Sec.~\ref{s:model-method}. We analyze the cluster formation,  structures and size distributions for dumbbells with asymmetric wetting properties in  Sec.~\ref{s:fluid1}.  In Sec.~\ref{s:fluid2} we present the results for dumbbells with asymmetric sizes.  Conclusions are given in Sec.~\ref{s:conc}.    

\section{Model and Methods}
\label{s:model-method}

\begin{figure}
\includegraphics[width=9cm]{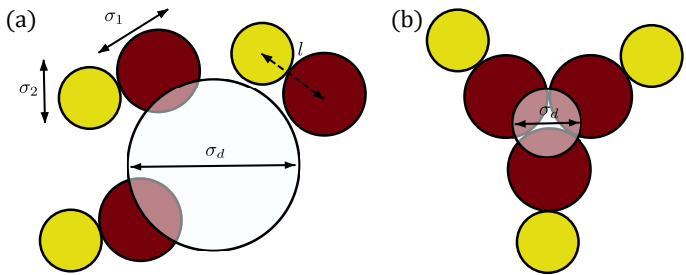}
\caption{Sketch of the model of colloidal dumbbells (bright yellow and dark red spheres) and droplets (white spheres). Shown are the diameters of colloidal species 1, $\sigma_{1}$, colloidal species 2, $\sigma_{2}$, and droplet $\sigma_{d}$. (a) In the initial stages the droplet captures the colloidal dumbbells. (b) The droplet has shrunk  and has pulled the dumbbells into a cluster. The competition between Yukawa repulsion and surface adsorption energies can lead to open cluster structures. }
\label{fig:skt}
\end{figure}

We simulate a ternary mixture of $N_\textrm{d}$ droplets of diameter $\sigma_{\textrm{d}}$ and $N_\textrm{c}$ colloidal dumbbells formed by two spherical colloids, labeled colloidal species 1 and colloidal species 2, of diameter $\sigma_{1}$ and $\sigma_{2}$ ($\sigma_{1}\geq \sigma_{2} $), respectively.  A sketch of the model is shown in Fig.~\ref{fig:skt}.
The colloids in each dumbbell are separated from each other by a distance $l$ that  fluctuates in the range of $\lambda\leq l\leq \lambda+\Delta $, where $\lambda=(\sigma_{1}+\sigma_{2})/2$. 

The total interaction energy is given by
\begin{eqnarray}
\dfrac{U}{k_\textrm{B}T} &=&\sum_{i<j}^{N_{c}} \phi_{11}\left ( \left | \mathbf{r}_{1i}-\mathbf{r}_{1j} \right | \right )+\sum_{i<j}^{N_{c}} \phi_{22}\left ( \left | \mathbf{r}_{2i}-\mathbf{r}_{2j} \right | \right )\nonumber \\
 &&+\sum_{i,j}^{N_{c}} \phi_{12}\left ( \left | \mathbf{r}_{1i}-\mathbf{r}_{2j} \right | \right )+\sum_{i}^{N_c} \sum_{j}^{N_{d}} \Phi _{1\textrm{d}}\left ( \left | \mathbf{r}_{1i}-\mathbf{R}_{j} \right | \right )\nonumber \\
&&+\sum_{i}^{N_c} \sum_{j}^{N_{d}} \Phi_{2\textrm{d}}\left ( \left | \mathbf{r}_{2i}-\mathbf{R}_{j} \right | \right )\nonumber \\
 &&+\sum_{i<j}^{N_{d}}  \Phi_{\textrm{dd}}\left ( \left |\mathbf{R}_{i}-\mathbf{R}_{j}  \right | \right ), 
 \label{eqn:total-energy}
\end{eqnarray}
where $k_\textrm{B}$ is the Boltzmann constant; $T$ is the temperature;  $\mathbf{r}_{1i}$ and $\mathbf{r}_{2i}$ are the center-of-mass coordinates of colloid 1 and colloid 2 in dumbbell $i$, respectively; $\mathbf{R}_{j}$ is the center-of-mass coordinate of droplet $j$; $\phi_{11}, \phi_{12}$ and $\phi_{22}$ are the colloid 1-colloid 1, colloid 1-colloid 2, and colloid 2-colloid 2 pair interactions, respectively;
$\Phi_{1\textrm{d}}$ and $\Phi_{2\textrm{d}}$ are the colloid 1-droplet, colloid 2-droplet pair interactions, respectively; and $\Phi_{\textrm{dd}}$ is the droplet-droplet pair interaction. 

The colloid-colloid pair interaction is composed of a short-ranged attractive square well and a longer-ranged repulsive Yukawa potential,
\begin{equation} 
  \phi_{11}(r)=\left \{ 
\begin{array}{ll}
 \infty  &  r<  \sigma_{1} \\
-  \epsilon_{\mathrm{SW}}
    &  \sigma_{1}  <r< \sigma_{1}+\Delta \\
  \epsilon_{\mathrm{Y}} \sigma_{1}\dfrac{e^{-\kappa \left ( r-\sigma_{1}  \right )}}{r}  & \textrm{otherwise,} 
    \end{array} \right . 
\label{eqn:phic1c1}
\end{equation}
\begin{equation} 
 \phi_{22}(r)=\left \{ 
\begin{array}{ll}
 \infty  &  r<  \sigma_{2} \\
-  \epsilon_{\mathrm{SW}}
    &  \sigma_{2}  <r< \sigma_{2}+\Delta \\
  \epsilon_{\mathrm{Y}}\sigma _{2}\dfrac{e^{-\kappa \left ( r-\sigma_{2}  \right )}}{r}  & \textrm{otherwise,} 
    \end{array} \right .  
\label{eqn:phic2c2}
\end{equation}
and
\begin{equation} 
 \phi_{12}(r)=\left \{ 
\begin{array}{ll}
 \infty  &  r<  \lambda  \\
-  \epsilon_{\mathrm{SW}}
    &  \lambda  <r< \lambda  +\Delta \\
  \epsilon_{\mathrm{Y}}\lambda\dfrac{e^{-\kappa \left ( r-\lambda   \right )}}{r}  & \textrm{otherwise,} 
    \end{array} \right . 
\label{eqn:phic1c2}
\end{equation}
where $r$ is the the center-center distance of particles. 
\begin{figure}
\includegraphics[width=4.25cm]{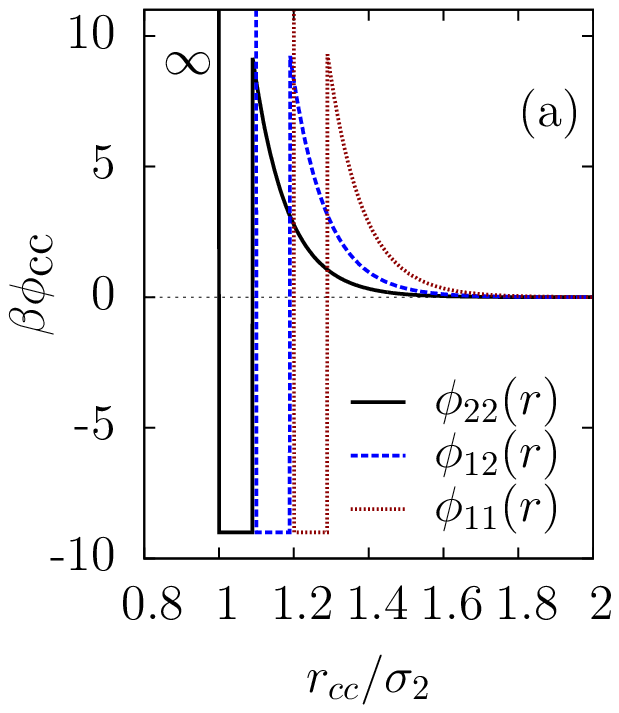}
\includegraphics[width=4.25cm]{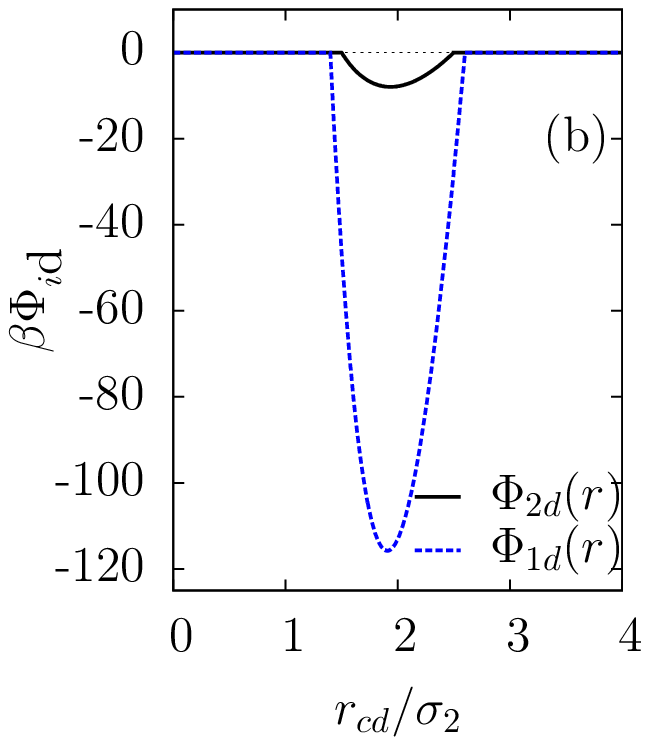}
\caption{Sketch of the pair interactions. (a) potentials between two colloidal particles with ${\epsilon_{\mathrm{SW}}=9k_{\textrm{B}}T}$, ${\Delta=0.09\sigma _{2}}$, $\epsilon_{\textrm{Y}}=24.6k_{\textrm{B}}T$ and  (b) colloid-droplet potential at ${\sigma_{d}\left(t\right)=4\sigma_{2}}$, with   $\sigma_1=1.2\sigma_2$.}
\label{fig:pot}
\end{figure} 

In Fig.~\ref{fig:pot}(a), the colloid-colloid interaction potentials are plotted against the separation for a given set of parameters used in the simulations. The parameters ${\epsilon_{\mathrm{SW}}=9k_{\textrm{B}}T}$, ${\Delta=0.09\sigma _{2}}$ are the depth and the width of a short-ranged attractive square well, respectively, while the parameter $\epsilon_{\textrm{Y}}=24.6k_{\textrm{B}}T$ controls the strength of long-ranged repulsive Yukawa interaction with inverse Debye length $\kappa\sigma_{2}=10$. 

A comparison between experimental quantities and simulation parameters can be found in Ref.~\cite{Mani2010}. In principle, the strength of the attractive interaction is chosen so that physical bonds between colloids  at the end of evaporation are irreversible. At the same time the repulsive barrier is chosen to be large enough to hinder spontaneous clustering. A wide range of simulation parameters satisfies the above conditions without qualitatively affecting the final results. The potential shape depicted in Fig.~\ref{fig:pot}(a) is similar to that employed by Mani \etal ~\cite{Mani2010}. However, differently from their systematic investigation of the repulsive parameters ($\epsilon_{\textrm{Y}}$,$\kappa$) on the stability of colloidal shells, we restrict our consideration to a fixed value of both attractive and repulsive parameters between colloids but vary colloid-droplet energies in order to address competing interactions.
 
The droplet-droplet pair interaction is a hard-sphere potential,
\begin{equation} 
 \Phi_{\textrm{dd}}(r)=\left \{ 
\begin{array}{ll}
 \infty  &  r<  \sigma_{\textrm{d}}+\sigma_{1}\\
  0 & \text{otherwise,} 
\end{array} \right . 
\label{eq:phidd}
\end{equation}
where the hard core droplet diameter $\sigma_{\textrm{d}}$ is added to the colloid diameter $\sigma _{1}$ such that no two droplets can share the same colloid (recall that $\sigma_{1}\geq\sigma_{2}$). 

The colloid-droplet interaction is taken to model the Pickering effect \cite{Pieranski1980}. Since the droplets shrink, their diameter is recorded as a function of time, that is, $\sigma_{d}(t)$ may be larger or smaller than that of the colloids. Hence, if $\sigma_{\textrm{d}} > \sigma_{i}$, the colloid-droplet adsorption energy is \cite{Ingmar2011}
\begin{equation} 
\Phi_{i\textrm{d}}(r)= \left \{ 
\begin{array}{ll}
 -   \gamma_{i}  \pi \sigma_{\textrm{d}} h &  \dfrac{\sigma_{\textrm{d}}-\sigma_{i}}{2}<r<  \dfrac{\sigma_{\textrm{d}}+\sigma_{i} }{2}\\
    0 & \textrm{otherwise},
\end{array} \right . 
\label{eqn:phicd1}
\end{equation}
and when $\sigma_{\textrm{d}} < \sigma_{i}$,
\begin{equation} 
\Phi_{i\textrm{d}}(r)= \left \{ 
\begin{array}{ll}
-    \gamma_{i}  \pi  \sigma_{\textrm{d}}^{2} &  r< \dfrac{\sigma_{i}-\sigma_{\textrm{d}}}{2}  \\
-    \gamma_{i}  \pi \sigma_{\textrm{d}} h &  \dfrac{\sigma_{i}-\sigma_{\textrm{d}}}{2}<r<  \dfrac{\sigma_{i}+\sigma_{\textrm{d}} }{2}\\ 0 & \textrm{otherwise,}
\end{array} \right . 
\label{eqn:phicd2}
\end{equation}
where $i=1,2$ labels the two colloidal species in each dumbbell, ${h=(\sigma_{i}/2-\sigma_{\textrm{d}}/2+r)(\sigma_{i}/2+\sigma_{\textrm{d}}/2-r)/(2r)}$  is the height of the spherical cap that results from the colloid-droplet intersection~\cite{Ingmar2011}, and the parameter $\gamma_{i}$ is the droplet-solvent interfacial tension used to control the strength of the colloid-droplet interaction. [See Fig.~\ref{fig:pot}(b) for an illustration of the colloid-droplet pair potential.] 

We introduce the energy ratio $k$ defined by 
\begin{equation}
k=\frac{\gamma _{2}}{\gamma _{1}},
\label{eqn:k}
\end{equation}
which characterizes the dissimilarity of the surface properties of the two colloidal species.

We define a bond between two colloidal spheres of type $i$ and $j$ when their distance is smaller than or equal to ${(\sigma_{i}+\sigma_{j})/2+\Delta}$, with $i,j=1,2$. A cluster is a group of colloidal particles connected with each other by a sequence of bonds. Hence, each cluster is characterized by both the number of bonds $n_b$ and the number of colloidal particles $n_c$ belonging to this cluster. A single dumbbell can be considered as a trivial cluster structure with $n_b=1, n_c=2$. These trivial clusters will be neglected in the following analysis.

We carry out Metropolis MC simulations in the \textit{NVT} ensemble. 
For a fixed set of parameters, statistical data are collected by running 30 independent simulations.  
In each run, a maximum displacement step of colloids ${d_{c}=0.01\sigma _{2}}$ and droplets ${d_{d}=d_{c}\sqrt{\sigma _{2}/\sigma _{\textrm{d}}}}$ ensures that Monte Carlo simulations are approximately equivalent to Brownian dynamics simulations~\cite{Sanz2010}. 

The total number of MC cycles per particle  is $10^{6}$, with $5\times 10^{5}$ MC cycles used to shrink the droplets at a fixed rate. This shrinking rate is chosen such that the droplet diameter vanishes after $5\times 10^{5}$ MC steps. 
Another $5\times 10^{5}$ MC cycles are used to equilibrate the  cluster configurations. As a test,  for  $k=0.1$ (open clusters), $k=0.5$ (intermediate clusters) and $k=1$ (closed clusters), we monitored  the total energy and the obtained number of clusters $N_{n_{c}}$ (composing of $n_c$ colloids  and $n_b$ bonds) for an additional $10^{6}$ cycles and found no changes in the results. 

Our kinetic MC simulation  uses sequential moves of individual particles
and  neglects the collective motion of particles in the cluster, i.e.\@ collective translational and rotational cluster moves are absent. 
Such collective modes of motion only play a role in dense colloidal suspensions of strongly interacting overdamped particles~\cite{Stephen2009,Stephen2011}. 

We did not attempt to reproduce the correct experimental time scale of  droplet evaporation, and the influence of colloid adsorption on the evaporation rate is neglected. The physical time corresponding to the MC time scale can be roughly estimated via the translational diffusion coefficient of clusters $D_{\textrm{cls}}$ defined by the Einstein relationship~\cite{Huitema1999}, 
\begin{equation}
\lim_{n\to\infty}\frac{\left\langle \triangle r_{\textrm{cls}}^{2}(n)\right\rangle }{n}=6D_{\textrm{cls}}\tau,
\end{equation}
where $n$ is the number of MC cycles and $\tau$ is the physical time per MC cycle. The Stokes-Einstein equation for diffusion of spherical particles is ${D_{\textrm{cls}}=\frac{k_{B}T}{3\pi\eta\sigma_{\textrm{cls}}}}$, with $\eta$ the viscosity of the solvent. Here $\left\langle \triangle r_{\textrm{cls}}^{2}(n)\right\rangle $ is the mean square displacement of the clusters after $n$ cycles, defined as 
\begin{equation}
\left\langle \triangle r_{\textrm{cls}}^{2}(n)\right\rangle =\dfrac{1}{N_{n_{c}}}{ \sum_{i=1}^{N_{n_{c}}}\triangle\mathbf{r}_{\textrm{cls},i}(n)\cdot\triangle\mathbf{r}_{\textrm{cls},i}(n)},
 \end{equation}
 where $N_{n_{c}}$ is the number of clusters with $n_c$ colloids and $\triangle\mathbf{r}_{\textrm{cls},i}(n)$ is the center-of-mass displacement of a cluster with $n_c$ colloids after $n$ cycles. In addition, the time required for a cluster to diffuse over its diameter $\sigma_{\textrm{cls}}$ is the so-called Brownian time scale $\tau_{B}$, given by ${\tau_{B}=\sigma_{\textrm{cls}}^{2}/D_{\textrm{cls}}}$ with the assumption that the diameter of the spherical cluster  ${\sigma_{\textrm{cls}}=\sqrt[3]{n_{c}}\sigma_2}$. Hence, we have
 \begin{equation}
 \dfrac{n\tau}{\tau_{B}}\simeq\frac{\left\langle \triangle r_{\textrm{cls}}^{2}(n)\right\rangle }{6\sqrt[3]{n_{c}^{2}}\sigma_{2}^{2}}
 \label{eqn:time}
  \end{equation}
  
  From Eq.~(\ref{eqn:time}) we derive an MC simulation time of about $10-20\tau_B$, depending on the number of colloids in the cluster.  As an example, for  clusters composed of ten colloids with diameter of $154\ \textrm{nm}$ in water ($\eta=1\ \textrm{mPa\ s}$) and at room temperature, we obtain a Brownian time $\tau_{B}\sim0.85\ \textrm{s}$. Compared to the time scales of experiments  that typically last tens of minutes~\cite{Wittemann2008,Ingmar2011}, our MC simulation time scales are much smaller. However, the validity of a similar model for a binary mixture of single colloidal particles and droplets has been demonstrated by qualitative and quantitative agreement between experimental and simulation results~\cite{Ingmar2011}.
  
 The simulations are performed in a cubic box with $N_c=250$ colloidal dumbbells with the packing fraction $\eta_c=0.01$ and $N_d=10-44$ droplets with packing fraction $\eta_d=0.1$. 
To initialize our simulation, we start by randomly distributing the colloidal dumbbells in the simulation box and with  random orientations. 
The initial distance between the colloid-1 and colloid-2  in a dumbbell is set smaller than $\lambda+\Delta$. 
In contrast, the initial distance between any two colloidal species that belongs to different dumbbells is larger than ${\sigma _{1}+\Delta}$. In this way, no two colloidal dumbbells bind together in the initial stage of the simulation. In addition, all colloids are located outside of the droplets. The initial droplet diameter is set to $8\sigma _{2}$ and shrunk at constant rate. 
  
\section{Results and Discussion}

\subsection{Asymmetric wetting properties and symmetric sizes}  
\label{s:fluid1}
\begin{figure}
\includegraphics[width=9cm]{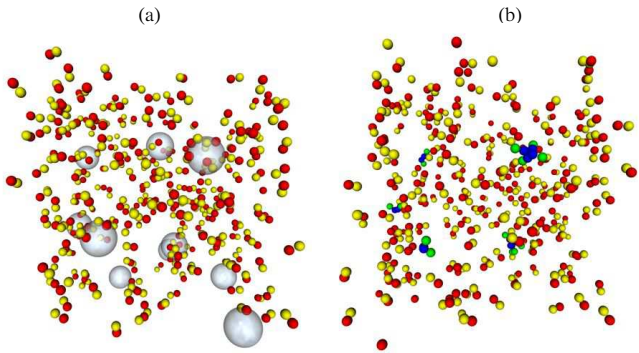}
\caption{Snapshots of the simulation for colloidal dumbbells with symmetric sizes and droplets at the energy ratio $k=0.1$. Results are shown at two different stages of the time evolution (a) after $2.5\times10^{5}$ MC cycles several colloidal dumbbells (bright yellow and dark red spheres) are trapped at the surface of the droplets (gray spheres) and (b) after $10^{6}$ MC cycles the stable clusters that are formed due to the droplets are composed of different colored colloids, that is, blue and green spheres represent colloidal species 1 and colloidal species 2, respectively. Open cluster structures with a compact core by colloid 1 and protruding arms by colloid 2 can be  observed.}
\label{fig:snap}
\end{figure}  

\begin{figure}
\includegraphics[width=9cm]{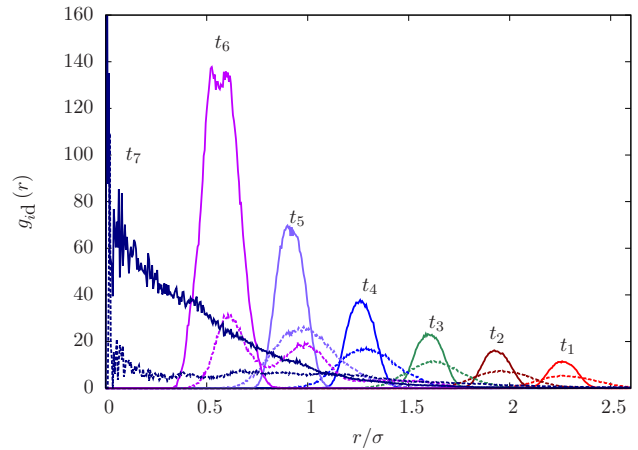}
\caption{Radial distribution functions, $g_{i\textrm{d}}(r)$ ($i=1,2$), for colloid 1-droplet (solid lines) and colloid 2-droplet (dashed lines) as a function of the scaled distance $r/\sigma$ at energy ratio $k=0.1$. Shown are results at different stages of the computer simulation. (See the notation in Table~\ref{tab:peak-position}.)}
\label{fig:fkt1}
\end{figure}
\begin{table}
\caption{Peak positions of the radial distribution functions $g_{1\textrm{d}}(r)$, $g_{2\textrm{d}}(r)$ and instantaneous droplet diameter at different stages of the time evolution for energy ratio $k=0.1$.}

\begin{tabular}{|c|c|c|c|c|}
\hline 
\multirow{2}{*}{$i$} & \multirow{2}{*}{$t_{i}$$\:\left(\times10^{5}\:\textrm{MC cycles}\right)$} & \multicolumn{2}{c|}{$\textrm{Peak positions}$$\:(r/\sigma)$} & \multirow{2}{*}{$\sigma_{\textrm{d}}(t)/\sigma$}\tabularnewline
\cline{3-4} 
 &  & $g_{1\textrm{d}}(r)$ & $g_{2\textrm{d}}(r)$ & \tabularnewline
\hline 
\hline 
$1$ & $2.0$ & $2.26$ & $2.27$ & $4.5$\tabularnewline
$2$ & $2.4$ & $1.92$ & $1.95$ & $3.9$\tabularnewline
$3$ & $2.8$ & $1.60$ & $1.62$ & $3.2$\tabularnewline
$4$ & $3.2$ & $1.26$ & $1.29$ & $2.5$\tabularnewline
$5$ & $3.6$ & $0.90$ & $0.97$ & $1.8$\tabularnewline
$6$ & $4.0$ & $0.55$ & %
\footnote{Peak 1%
}$0.60\quad1.00$%
\footnote{Peak 2%
} & $1.1$\tabularnewline
$7$ & $5.0$ & $\parallel$ & $\parallel$%
\footnote{Undefined value%
} & $0.0$\tabularnewline
\hline 
\end{tabular} 
\label{tab:peak-position}
\end{table}

We first study dumbbells built of colloids with equal diameter  $\sigma_{1}=\sigma_{2}\equiv\sigma$ and different wetting properties.  
The parameter $\gamma _{1}$ is fixed to $100k_{\textrm{B}}T/\sigma^{2}$, while the parameter $\gamma _{2}$ is varied from $10k_{\textrm{B}}T/\sigma^{2}$ to $100k_{\textrm{B}}T/\sigma^{2}$. As a consequence, the energy ratio, Eq.~\ref{eqn:k}, ranges from $ k=0.1-1 $. In the special case of $k=1$,  colloids 1 and 2 are identical. 

Figure~\ref{fig:snap} shows snapshots at two different stages of the simulation for the energy ratio $k=0.1$. After $2.5\times10^{5}$ MC cycles [see Fig.~\ref{fig:snap}(a)] colloidal dumbbells are captured at the droplet surface. Figure ~\ref{fig:snap}(b) shows the final cluster configurations obtained after $10^{6}$ cycles. Only clusters that are stable against thermal fluctuations survived and are considered for analysis.

We analyze how colloidal dumbbells are captured by the droplet surface by means of the radial distribution functions of colloid 1-droplet, $g_{1\textrm{d}}(r)$, and colloid 2-droplet, $g_{2\textrm{d}}(r)$, defined explicitly as $g_{i\textrm{d}}(r)=\frac{dn_{i\textrm{d}}(r)}{4\pi r^{2}dr\rho_{\textrm{d}}}$ with $dn_{i\textrm{d}}(r)$ the number of droplets between distances $r$ and $r+dr$ from a colloid of species $i$ ($i=1,2$) and $\rho_{\textrm{d}}$ the average number density of droplets.   
In Fig.~\ref{fig:fkt1}, we consider different stages of the time evolution ranging from $t_1$ to $t_7$ (see Table~\ref{tab:peak-position} for an explanation of the symbols).  Between times $t_1$ and $t_6$ the function $g_{1\textrm{d}}(r)$ (solid lines) shows only a single peak. 
For example, at $t_1=2\times10^{5}$ MC, $g_{1\textrm{d}}(r)$ has a peak at $r\simeq 2.25\sigma$ corresponding to the instantaneous droplet radius $\sigma_{\textrm{d}}(t)/2$. The peak is due to colloid-1 spheres trapped at the droplet surface. 
The droplet radius decreases continuously during the modeled evaporation. As a result, the peak position of $\sigma_{\textrm{d}}(t)/2$ shifts continuously towards smaller  distances. Moreover, since the number of trapped type-1 colloids onto the droplet surface can increase during the movement of  particles, the peak height of $g_{1\textrm{d}}\left ( r \right )$  increases with MC time. 
Finally, after $t=t_7$ ($5\times10^{5}$ MC cycles) the droplets vanish completely [$\sigma _{\textrm{d}}(t)=0$] and as a result  $g_{1\textrm{d}}(r)$ stops changing. 

A similar trend can be observed in the radial distribution function $g_{2\textrm{d}}(r)$ (dashed lines in Fig.~\ref{fig:fkt1}). However, $g_{2\textrm{d}}(r)$ has two distinct peaks at $t_6=4\times10^{5}$ MC cycles. Table~\ref{tab:peak-position} lists peak positions of $g_{1\textrm{d}}(r)$, $g_{2\textrm{d}}(r)$ and instantaneous droplet diameter $\sigma_{\textrm{d}}(t)$ with respect to the time evolution of the system for $k=0.1$. 
In addition, for $k=0.1$ the peak height of $g_{1\textrm{d}}\left ( r \right )$ is always much larger than that of $g_{2\textrm{d}}\left ( r \right )$ at the same time. This means that there exists a higher probability of finding type-1 colloids  than finding type-2 colloids on the droplet surface.

\begin{figure}
\includegraphics[width=4.25cm]{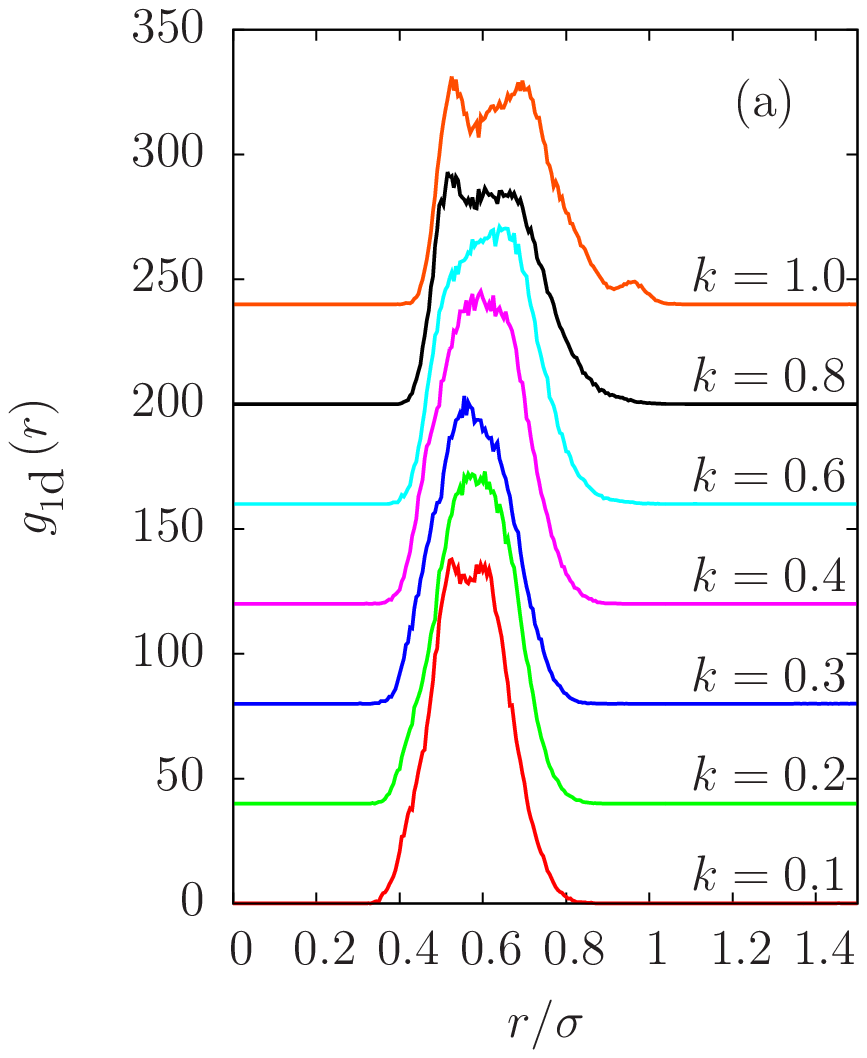}
\includegraphics[width=4.25cm]{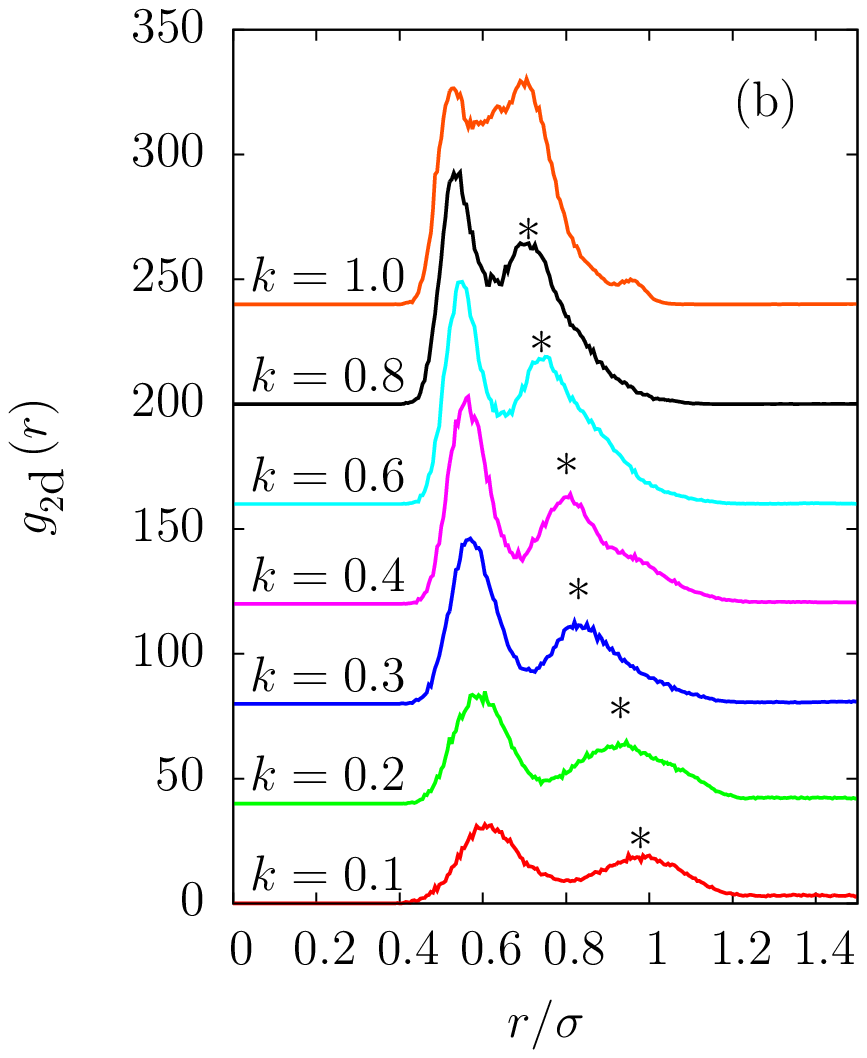}
\caption{Colloid 1-droplet (a) and colloid 2-droplet (b) radial distribution functions, $g_{1\textrm{d}}(r)$ and $g_{2\textrm{d}}(r)$, respectively, as a function of the scaled distance $r/\sigma$ after $t=t_6$ ($4.0\times 10^{5}$ MC cycles). Results are shown for different energy ratios $k$. An asterisk is used as a guide to the eyes to trace the shift of the second peak. Curves are shifted upwards by 40 units for clarity.}
\label{fig:fkt2}
\end{figure}

Figure~\ref{fig:fkt2} shows results for $g_{1\textrm{d}}(r)$ and $g_{2\textrm{d}}(r)$ at different energy ratios $k$ after $4.0\times 10^{5}$ MC cycles. 
As shown in Fig.~\ref{fig:fkt2}(a), $g_{1\textrm{d}}(r)$ has a peak at $r\simeq \sigma _{\textrm{d}}(t)/2$ that is independent of the value of $k$. At the same time, $g_{2\textrm{d}}(r)$ [Fig.~\ref{fig:fkt2}(b)] exhibits two distinct peaks, the first peak at a position coinciding with the peak of $g_{1\textrm{d}}(r)$, and the second peak (marked by an asterisk), which shifts towards the first peak with increasing $k$. 
Colloids trapped on the droplet surface feel the Yukawa repulsive interaction, thermal fluctuation and adsorption interaction between colloids and droplets $\Phi _{i\textrm{d}}$, $i=1,2$. 
For a given colloid 1-droplet interaction $\gamma _{1}=100k_{\textrm{B}}T/\sigma^{2}$, whose magnitude is much larger than the Yukawa repulsive interaction and thermal energy, the colloid-1 spheres cannot overcome the energy barrier to escape from the droplet surface. Meanwhile, for $k=0.1$ ($\gamma _{2}=10k_{\textrm{B}}T/\sigma^{2}$) the trapped colloid-2-droplet interaction may be comparable to the Yukawa repulsive interaction and thermal energy. This leads to some colloid-2 to be separated from each other and/or released from the droplet surface, forming the second peak at a distance larger than $\sigma _{\textrm{d}}(t)/2$. 
When the energy ratio $k$ increases, the binding energy between trapped colloid-2 and droplets becomes stronger, which results in an increase of the probability of finding the colloid-2 at a shorter radial distance from the droplet.  
Finally, for $k=1$, all of trapped colloid-1 and colloid-2 are strongly localized on the droplet surface, signalled by a single peak with a broader width (see in Fig.~\ref{fig:fkt2}).           

\begin{figure}
\includegraphics[width=8.5cm]{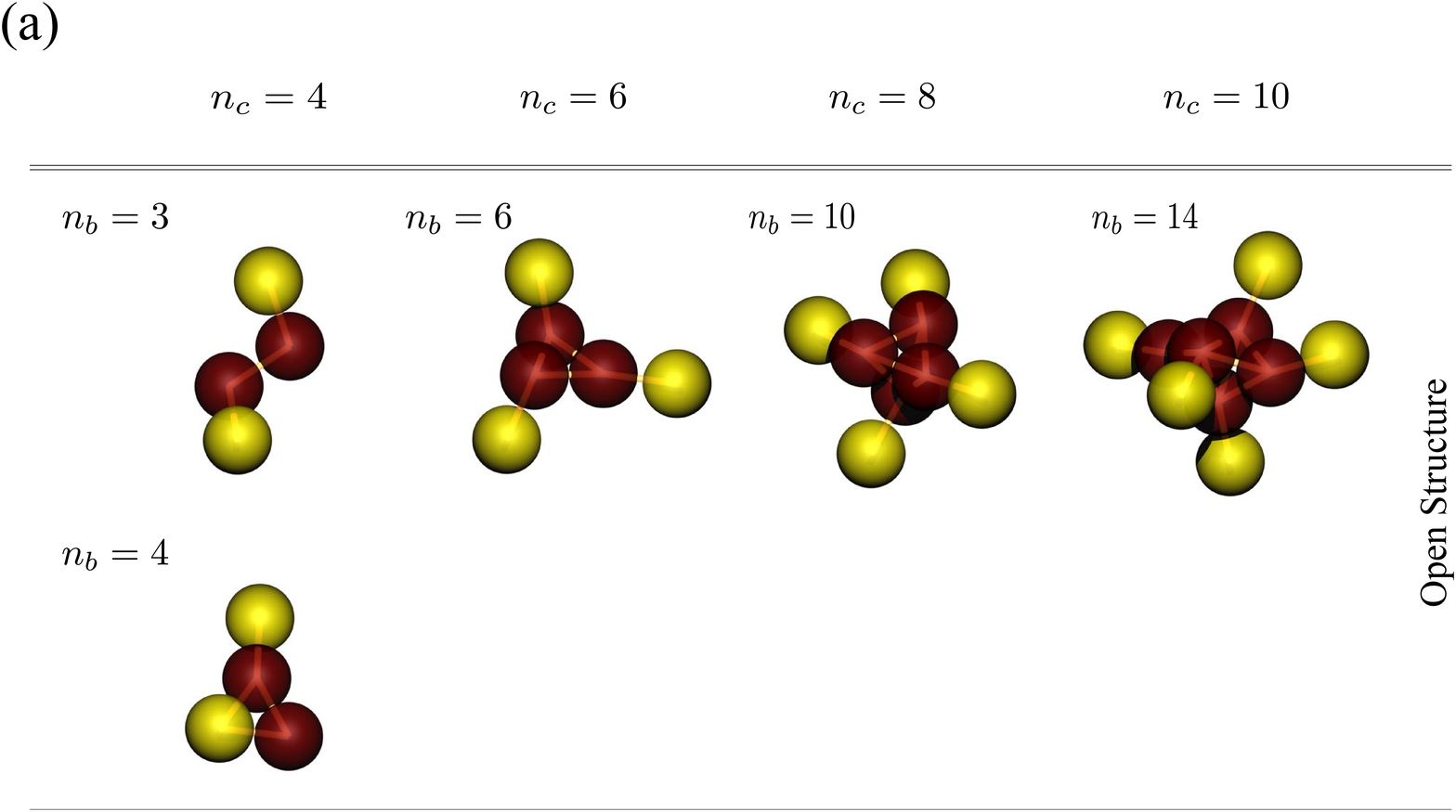}
\includegraphics[width=8.5cm]{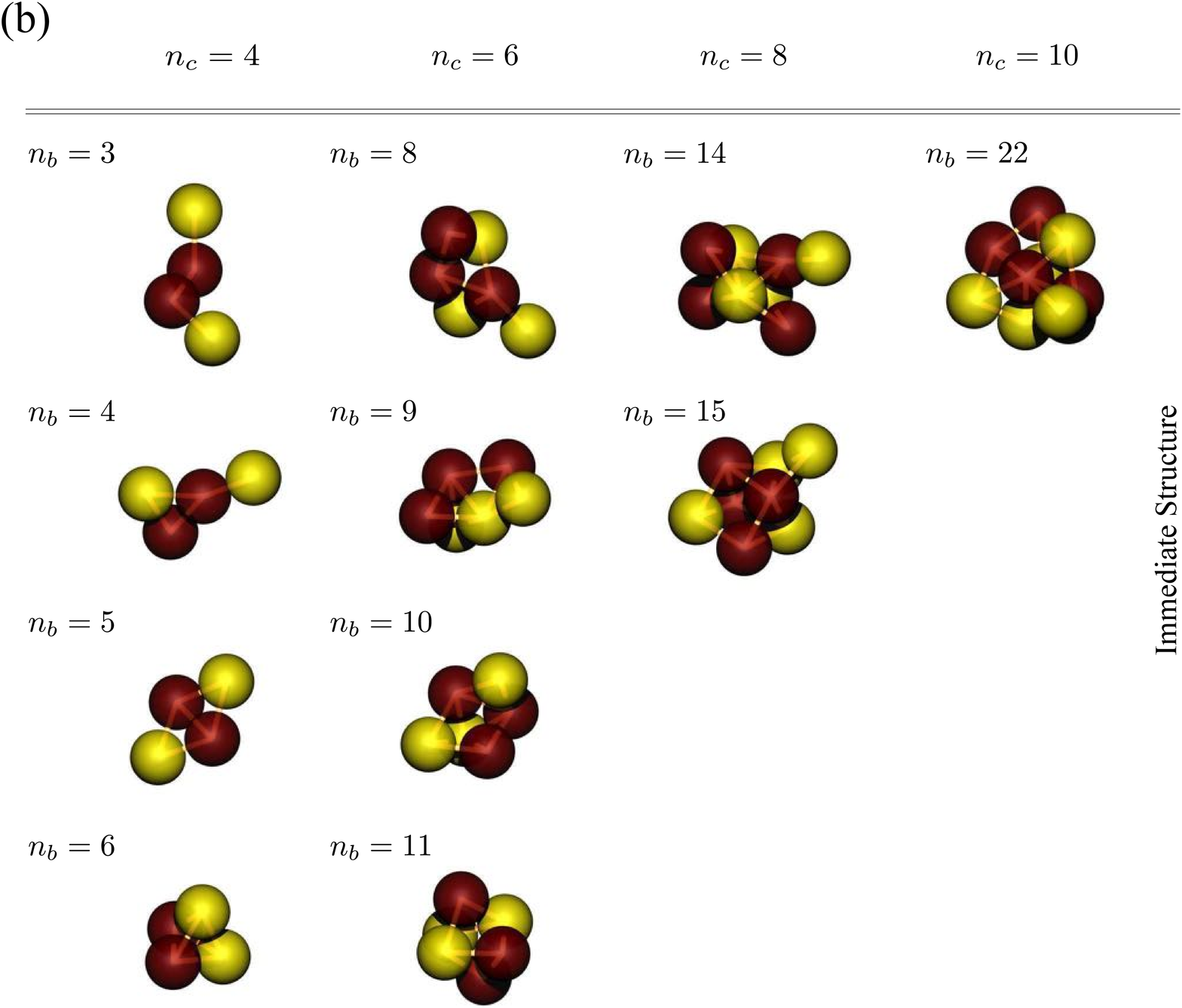}
\includegraphics[width=8.5cm]{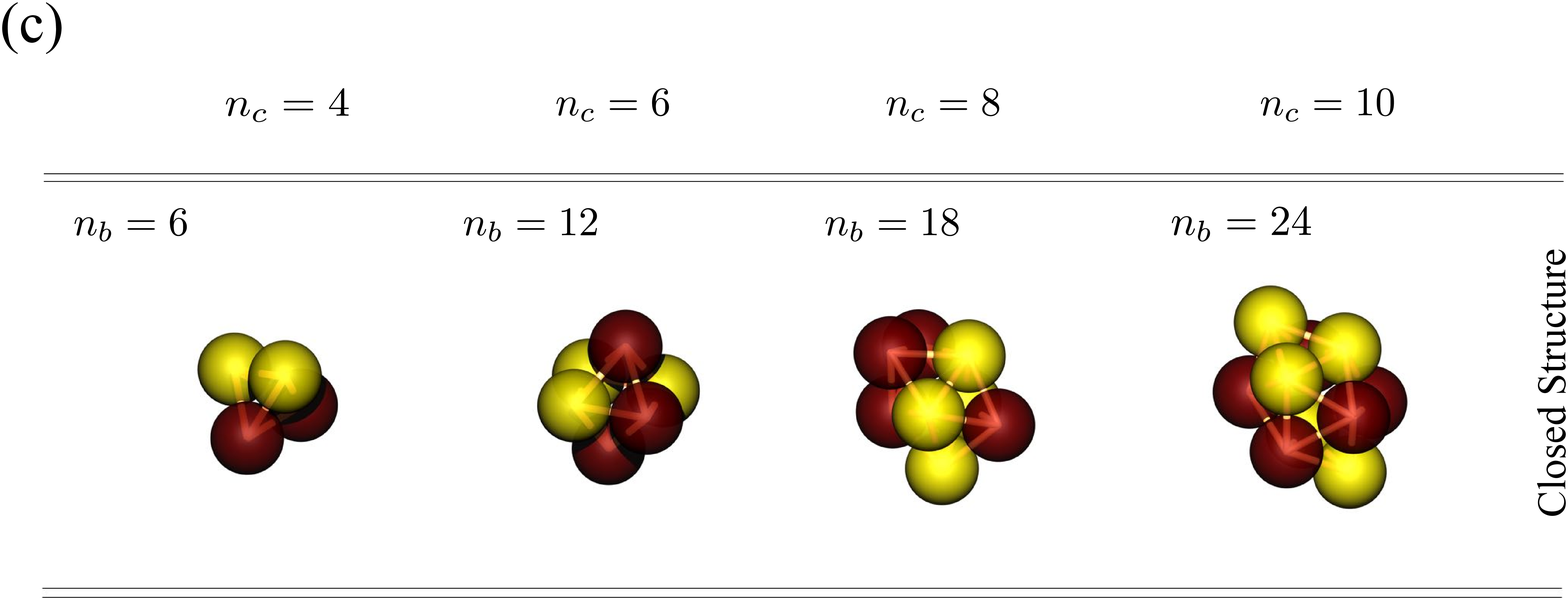}
\caption{Typical cluster structures found in simulations for (a) $k=0.1$, (b) $k=0.5$ and (c) $k=1$ at the final stage of the simulations. The red and yellow colored spheres represent colloid-1 and colloid-2 spheres in each dumbbell, respectively. For each cluster with the same number of constituent colloids the bond number $n_b$ is used to distinguish whether a cluster is open or closed structure. The wireframe connecting the colloid centers represent the bond skeleton.}
\label{fig:cluster}
\end{figure}

Examples of the obtained cluster structures are shown in Fig.~\ref{fig:cluster}(a) for $k=0.1$, Fig.~\ref{fig:cluster}(b) for $k=0.5$ and Fig.~\ref{fig:cluster}(c) for $k=1$. 
Clusters with colloid numbers between $n_c=4$ and $n_c=10$ are found. 
For the same number of constituent colloids $n_c$, clusters can have several distinct structures (isomers)~\cite{Ingmar2011,Bo2013}. 
It is convenient to use the bond-number $n_b$ as an indicator for the compactness of clusters. For a given value of $n_c$, the smaller the bond number $n_b$ is, the more open the structure is. 
As shown in Fig.~\ref{fig:cluster}(a),  open structures are obtained for $k=0.1$. In these isomers, the colloids of type 1 arrange themselves into symmetric structures, i.e, doublet, triplet, tetrahedron and triangular dipyramid. 
Increasing the energy ratio $k$, a larger number of isomers with different bond number $n_b$ are found. 
For example, for $n_c=4$ [Fig.~\ref{fig:cluster}(b)] we find four different isomers with $n_b$ ranging from $3$ to $6$, corresponding to a transition from string-like clusters to more compact structures. 
Finally, for the special case $k=1$ the two colloidal species are identical and we find compact isomers with the largest $n_b$ [Fig.~\ref{fig:cluster}(c)] such as $n_c=4,n_b=6$ (tetrahedron); $n_c=6,n_b=12$ (octahedron); $n_c=8,n_b=18$ (snub disphenoid); $n_c=10,n_b=22$ (gyreoelongate square dipyramid). These structures are similar to the one-component structures that minimize the second moment of the mass distribution \cite{Manoharan2003,Wittemann2010}.  

\begin{figure*}
\includegraphics[width=5.9cm]{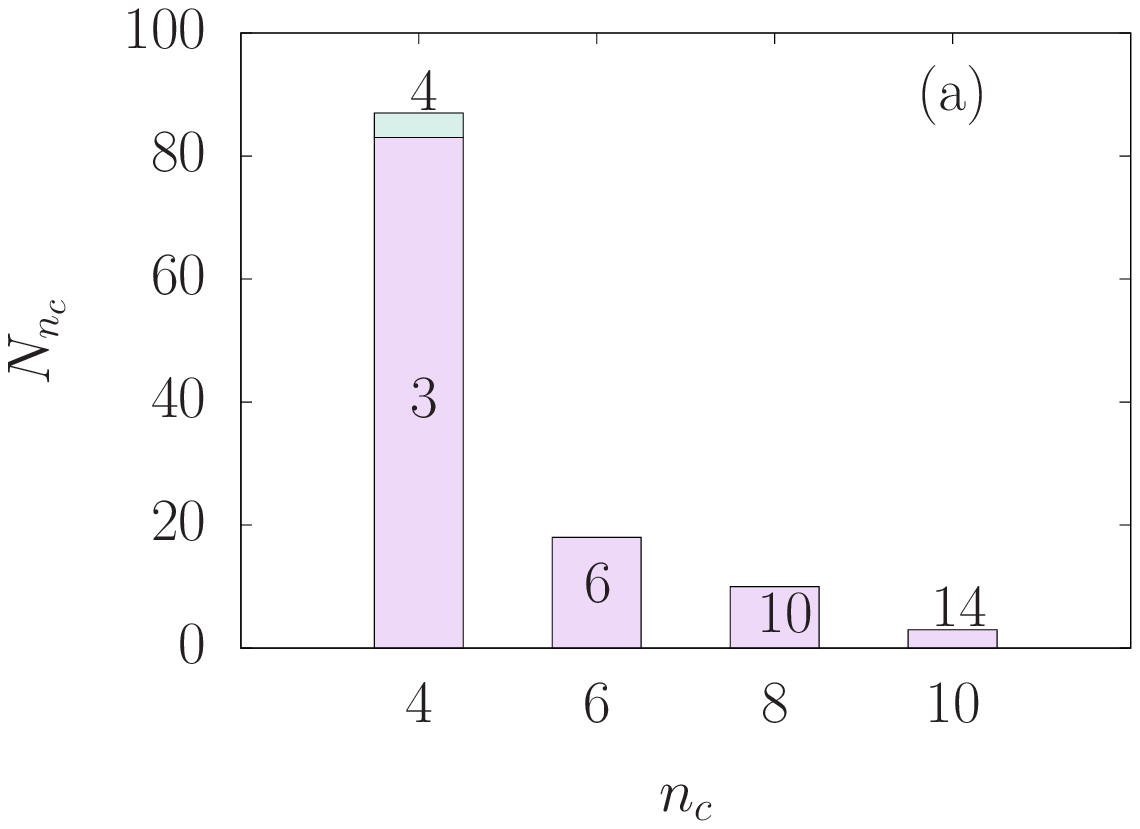}
\hspace{-0.3cm}
\includegraphics[width=5.9cm]{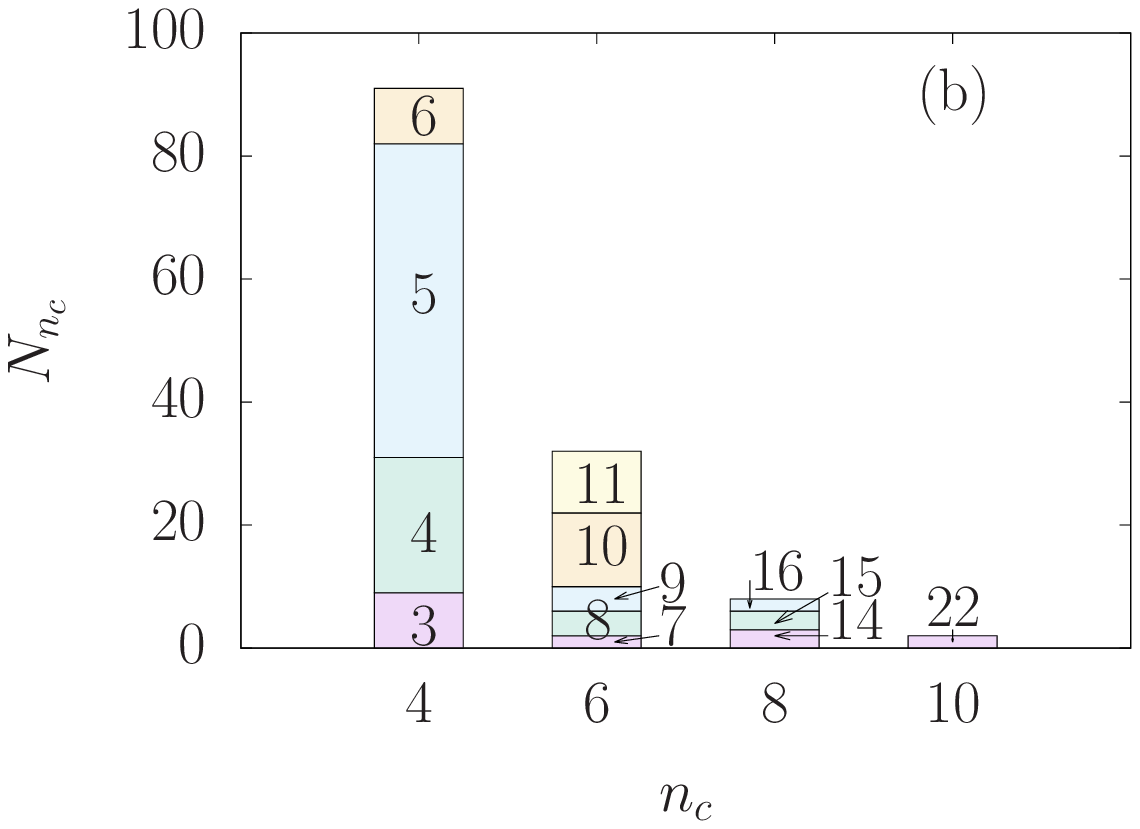}
\hspace{-0.3cm}
\includegraphics[width=5.9cm]{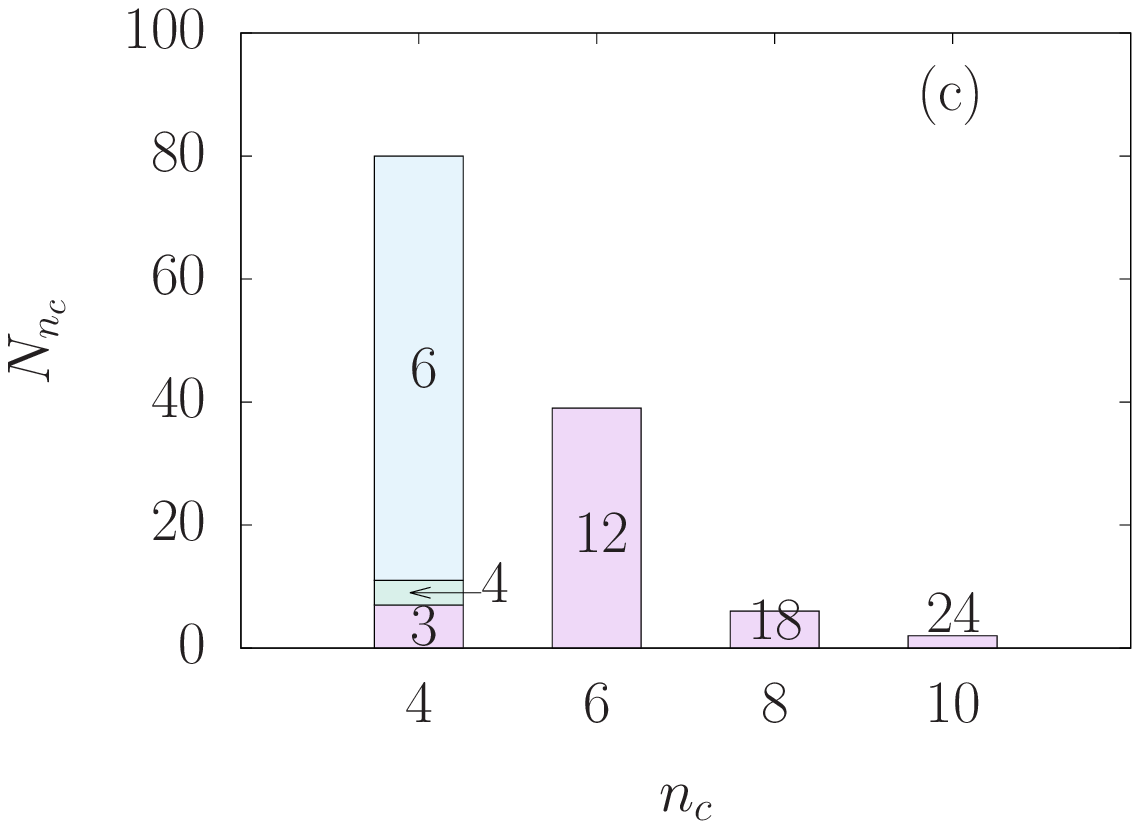}
\caption{Distribution of the number of clusters $N_{n_{c}}$ as a function of the number of colloids $n_c$ in the cluster in the final stage of simulation. Results are shown for different energy ratios (a) $k=0.1$, (b) $k=0.5$ and (c) $k=1$ at $\sigma _{1}=\sigma _{2}\equiv\sigma$ and $\gamma_1=100k_{\textrm{B}}T/\sigma^{2}$. The colored region is labeled with the bond number $n_b$. }
 \label{fig:hist1}
 \end{figure*}

Figure~\ref{fig:hist1} shows a stacked histogram of the number of clusters $N_{n_{c}}$ with $n_c$ colloids. The height of each differently colored bar is proportional to the number of clusters with the bond number $n_b$. For small value of $k$, a large fraction of clusters has an open structure, while for $k=1$ almost all cluster have closed structure and for $k=0.5$ a variety of intermediate structures can be found. These observations are in good agreement with our results for colloid-droplet radial distribution functions, as discussed above. 
  
\subsection{Symmetric wetting properties and asymmetric sizes}  
\label{s:fluid2}
We next investigate the cluster formation of colloidal dumbbells built with spheres of different diameters, ${\sigma_{1}=1.5\sigma_{2}}$ and $\sigma_{1}=2.0\sigma_{2}$ but equal wetting properties, which are obtained by the interfacial tension $\gamma_1=\gamma_2\equiv\gamma$. 
We investigate the setting values  $\gamma=$10, 40 and  $100k_{\textrm{B}}T/\sigma _{2}^{2}$. We note that a size asymmetry  between the  colloids forming the dumbbells causes an asymmetry in colloid-droplet adsorption energies  [Eqs.~(\ref{eqn:phicd1}) and~(\ref{eqn:phicd2})].  For this reason, the structures found in this case are the same as those shown in Fig.~\ref{fig:cluster} for asymmetric wetting properties.

\begin{figure*}
\begin{tabular}{ccc}
\includegraphics[width=5.9cm]{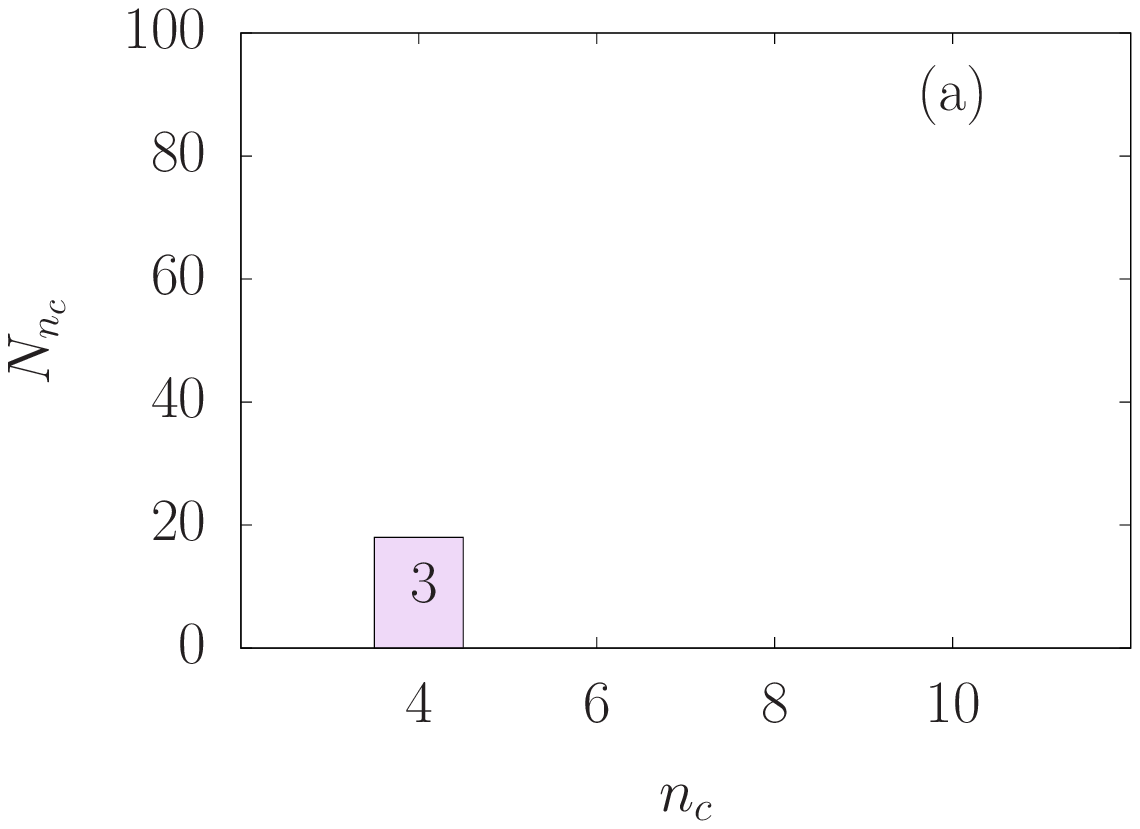}&
\hspace{-0.3cm}
\includegraphics[width=5.9cm]{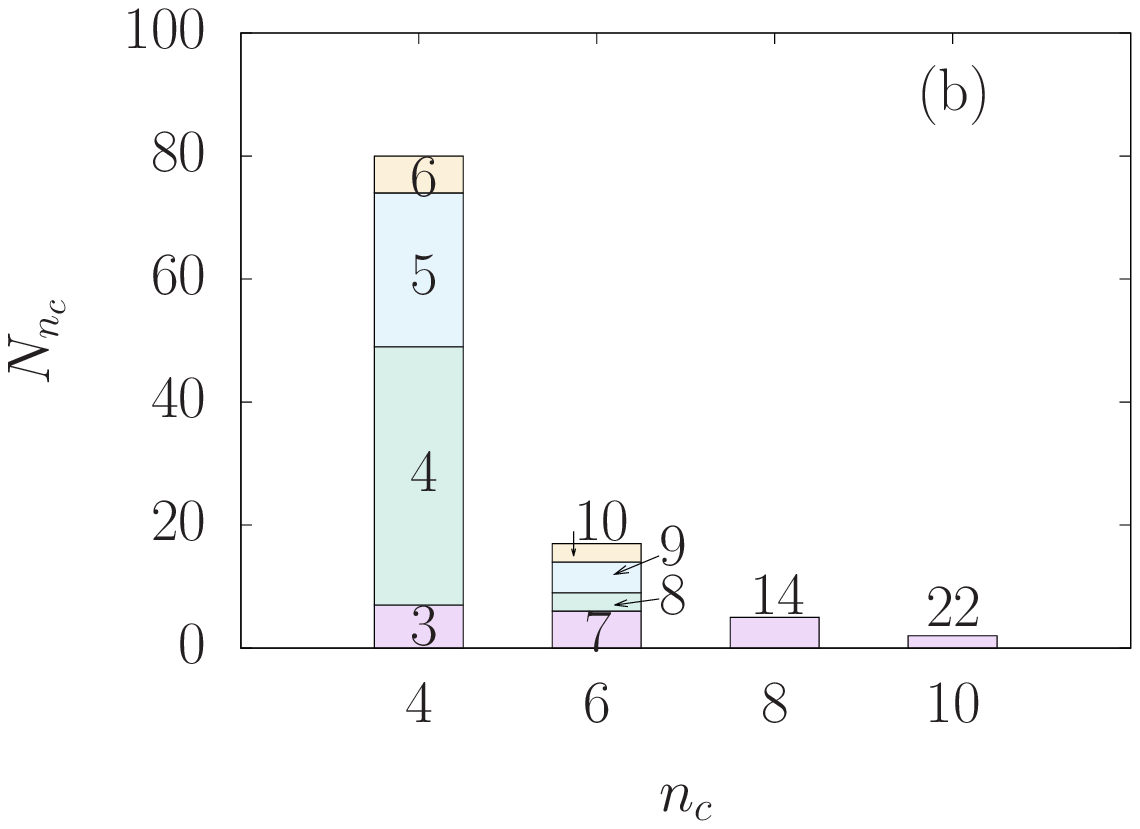}&
\hspace{-0.3cm}
\includegraphics[width=5.9cm]{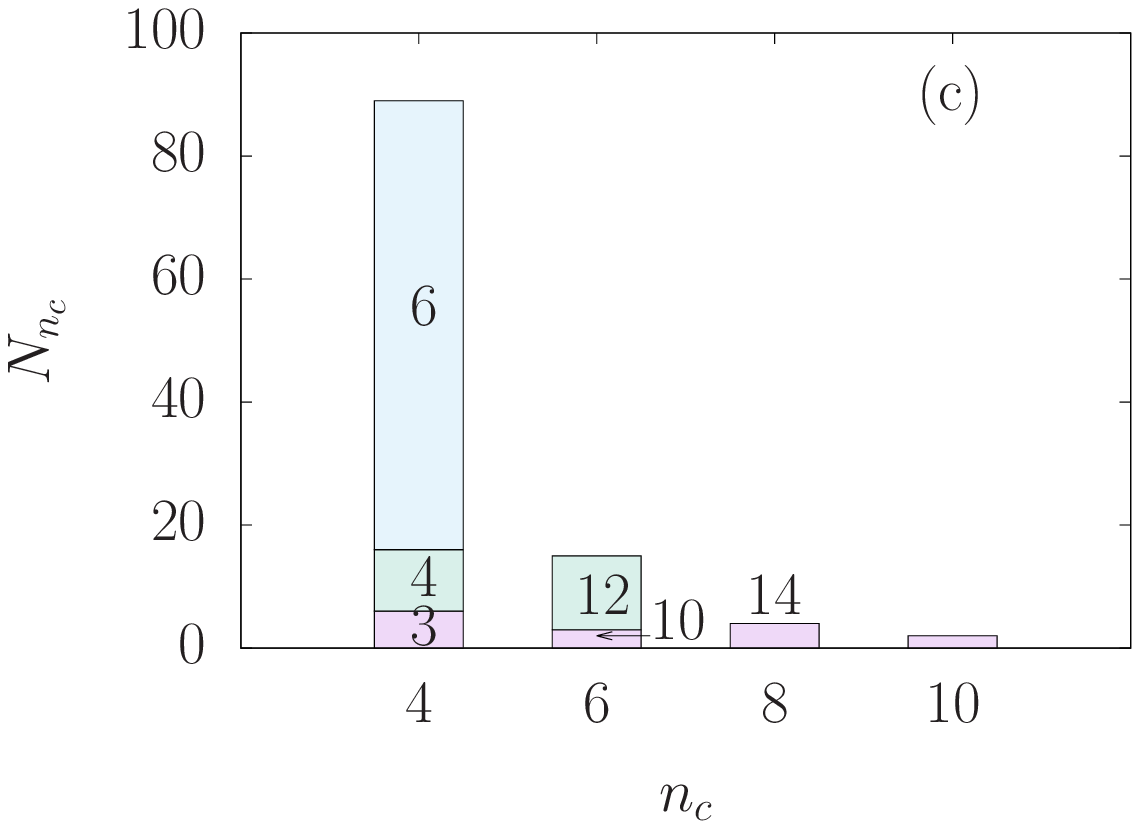}\\
\includegraphics[width=5.9cm]{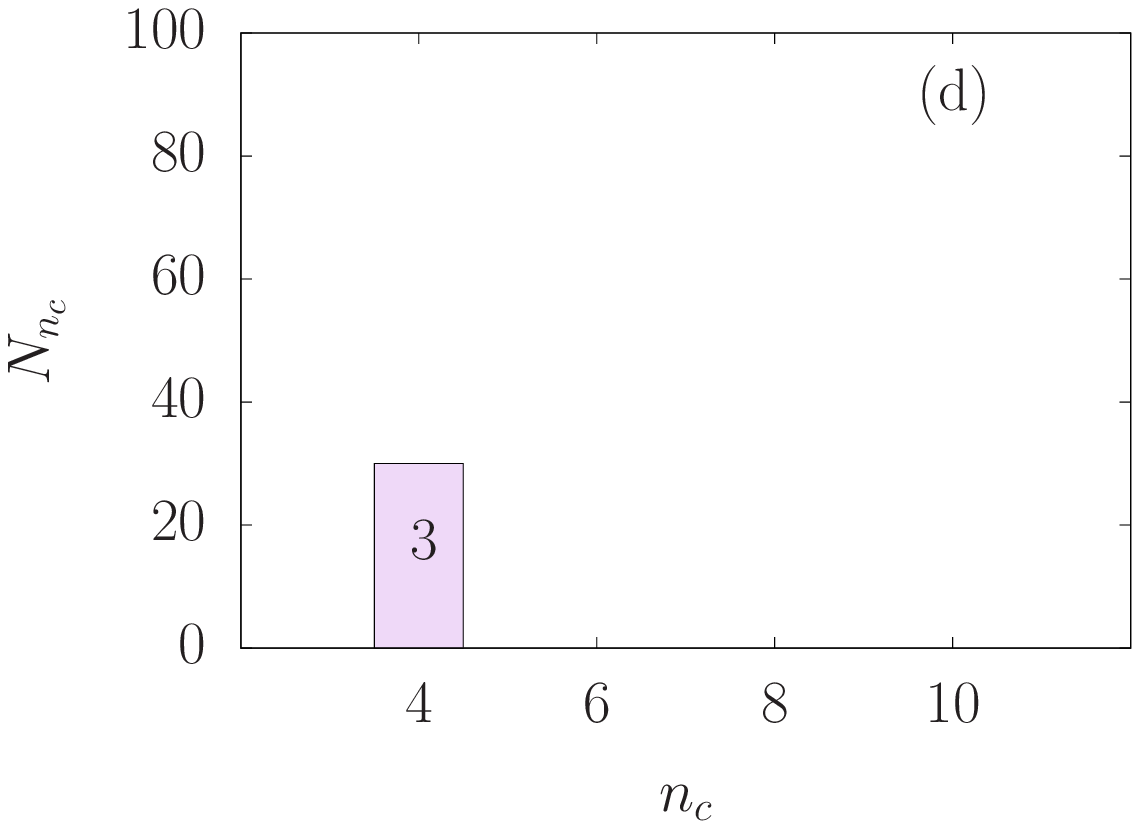}&
\hspace{-0.3cm}
\includegraphics[width=5.9cm]{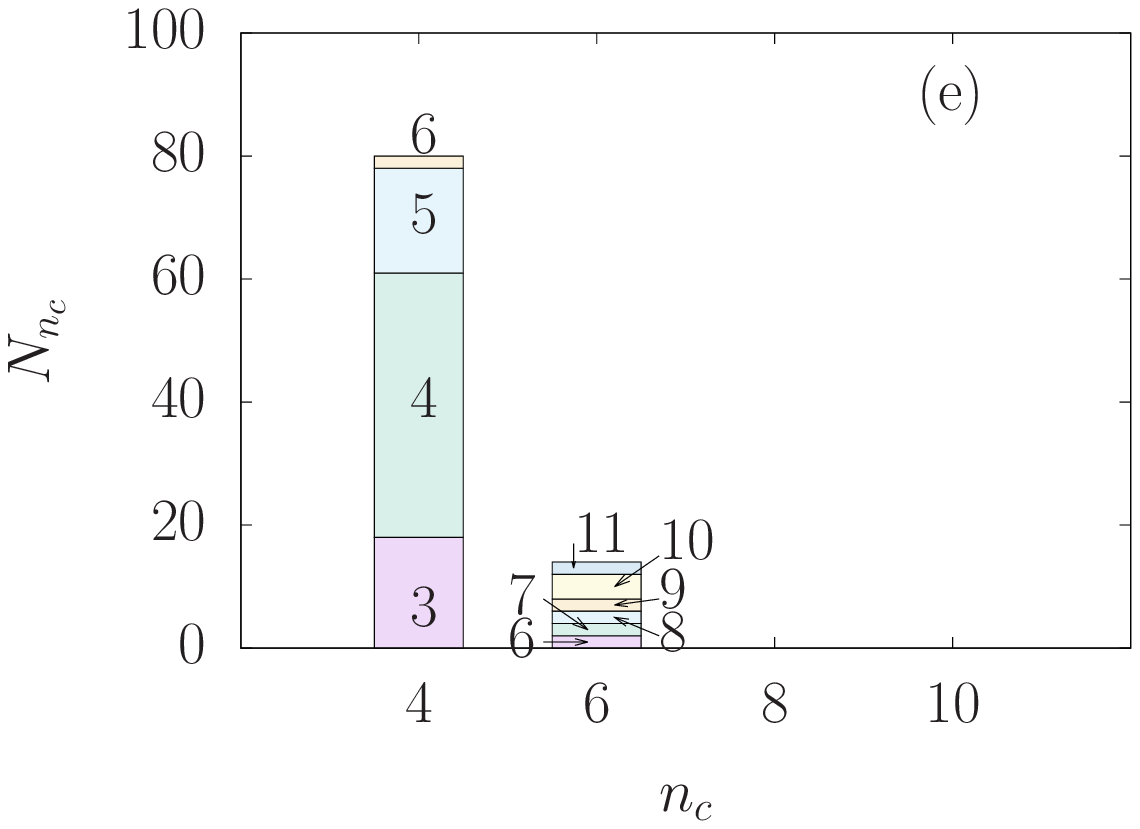}&
\hspace{-0.3cm}
\includegraphics[width=5.9cm]{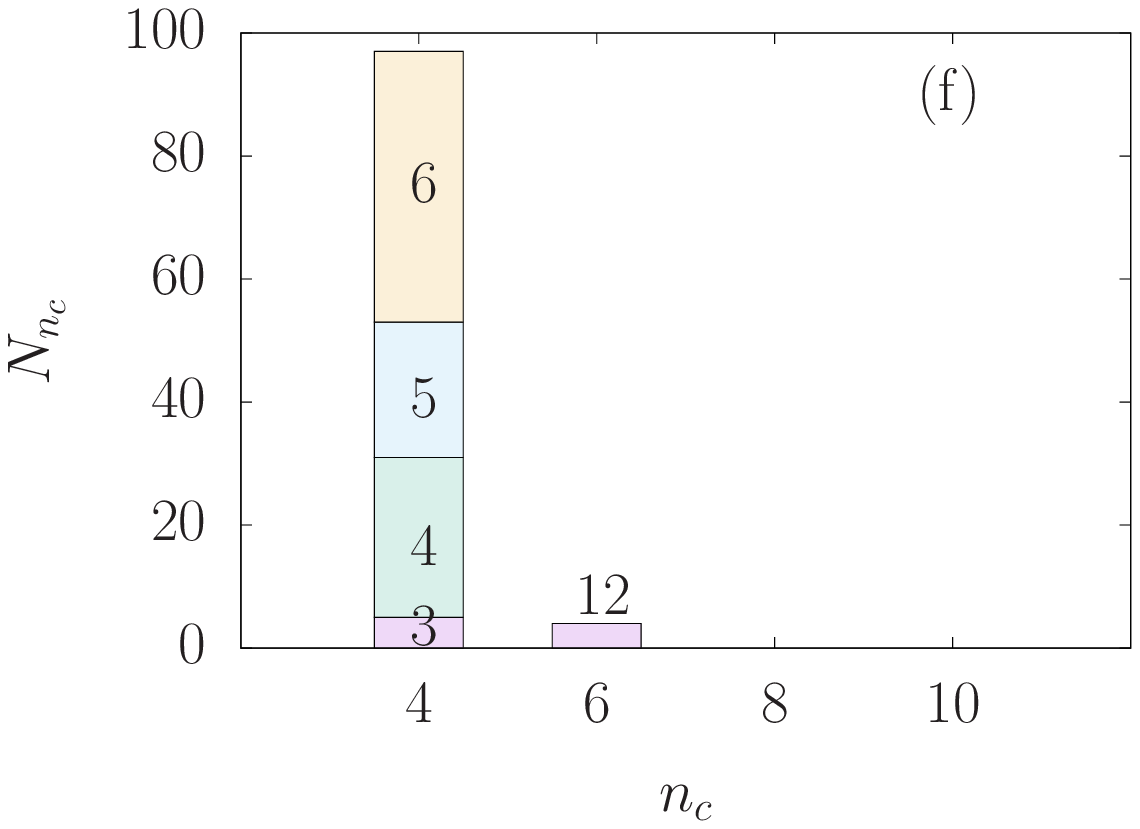}\\
\end{tabular}
\caption{Distribution of the number of clusters $N_{n_{c}}$ as a function of the number of colloids $n_c$ in the cluster in the final stage of simulation. Results are for different interfacial tensions as indicated (a), (d) $\gamma=10k_{B}T/\sigma_{2}^{2}$; (b), (e) $\gamma=40k_{B}T/\sigma _{2}^{2}$ and (c), (f) $\gamma=100k_{B}T/\sigma _{2}^{2}$ at $\sigma_{1}=1.5\sigma_{2}$ and $\sigma_{1}=2.0\sigma_{2}$, respectively. The numerical label in differently colored regions indicates the bond number $n_b$.}
\label{fig:hist2}
\end{figure*}

We analyze the size distribution of the clusters. Figure~\ref{fig:hist2} shows stacked histograms of the number of clusters $N_{n_{c}}$ with $n_c$ colloids for different values of  $\gamma$ and two different size ratios. 
For the case $\sigma_{1}=1.5\sigma_{2}$ and $\gamma=10k_{\textrm{B}}T/\sigma _{2}^{2}$  [Fig.~\ref{fig:hist2}(a) and Fig.~\ref{fig:hist2}(d)], all clusters have open structures with  $n_b=3$. 
In addition, we do not find cluster with a high  $n_c$ [Fig.~\ref{fig:hist2}(a) and (d)] because the Yukawa repulsion and the thermal fluctuations dominate over the adsorption energy between colloids and droplets that keeps the colloids in a compact arrangement. 
On the other hand, in the case of $\gamma=40k_{\textrm{B}}T/\sigma _{2}^{2}$ [Fig.~\ref{fig:hist2}(b)] we observe many clusters of bond numbers $n_b$ in the range $3-6$, corresponding to intermediate structures.
Finally, when  $\gamma=100k_{\textrm{B}}T/\sigma _{2}^{2}$, the adsorption energy between colloids and droplets is much larger than the total repulsive energy. Therefore, we observe mostly closed structures [Fig.~\ref{fig:hist2}(c)]. 

At a larger size asymmetry of $\sigma_{1}=2.0\sigma_{2}$, but at the same interfacial tension $\gamma=40, 100\, k_{\textrm{B}}T/\sigma _{2}^{2}$ [Fig.~\ref{fig:hist2}(b),(f) and~\ref{fig:hist2}(c),(f)] we observe a decrease of the number of large clusters, while the yield of smaller clusters increases. 

\section{Summary and Conclusions}
\label{s:conc}
We investigated the cluster formation process of a mixture of colloidal dumbbells and droplets via emulsion droplet evaporation using Metropolis-based kinetic Monte Carlo simulations. The short-ranged attraction between colloids has a potential well depth of $9k_{\textrm{B}}T$ in order to ensure that neither dumbbells nor clusters are likely to break apart due to thermal fluctuations. 
In addition, the height of the repulsive barrier between colloids is about $9k_{\textrm{B}}T$, which is a large enough value to avoid spontaneous formation of clusters. 
The droplet-droplet interaction is a hard-sphere repulsion with an effective hard-sphere diameter chosen so that any two droplets cannot merge. The adsorption interaction between colloids and droplets has a minimum at the droplet surface to model the Pickering effect. In experiments, this energy has values up to millions of $k_{\textrm{B}}T$, depending on the contact angle, interfacial tension and particle size~\cite{Aveyard2003}. In our simulations, however, we limited the colloid-droplet adsorption energy below $100k_{\textrm{B}}T$, and the contact angle at a planar interface is $90^\circ$.  

In the dumbbell system with symmetric sizes, the colloid 1-droplet adsorption energy is kept at a fixed value of nearly $100k_{\textrm{B}}T$, while the colloid 2-droplet adsorption energy is controlled by changing the interfacial tension. Droplet-colloid radial distribution functions indicate that both colloid 1 and colloid-2 spheres can be captured and freely diffuse on the droplet surface. Choosing a smaller colloid 2-droplet energy leads to an increase of the probability of colloid 2 detachment from the droplet surface. In agreement with typical cluster structures in the final stage of simulation we found that clusters with the same number of constituent colloids can produce a variety of different isomers. The bond number was used to assess whether an isomer is open or closed.  Histograms show that a larger fraction of open isomers can be obtained by decreasing the colloid 2-droplet adsorption energy. 

Similar results were obtained in the asymmetric dumbbell system. Whether open, intermediate or closed structures are formed, strongly depends on the interfacial tension of both colloid 1 and colloid 2 and their relative sizes. 
This results from competing Yukawa repulsion, colloid-droplet adsorption interactions and thermal fluctuations. However, choosing a larger size of colloid 1 compared to colloid 2 could lead to a decrease in the number of large clusters. 

Although closed structures have been reported in many studies of the assembly of single component spheres~\cite{Manoharan2003, Wittemann2010,Ingmar2011}, the open and intermediate structures found here have not yet been observed in experiments. The repulsive energy between the colloids can be controlled experimentally by tuning pH, concentration of salt, and composition of the solution~\cite{Mani2010}, while the adsorption energy can be controlled by colloid diameter and wettability~\cite{Bernard2008}. Therefore, our result could be useful to guide experimental work for preparing increasingly complex building blocks for the assembly of nanostructured materials.

\begin{acknowledgments}
This research was supported by a PhD grant of the Vietnamese Government Scholarship Program (Project 911).
\end{acknowledgments}

\bibliographystyle{apsrev}
\bibliography{refs}

\end{document}